\documentstyle[eqsecnum,preprint,aps,epsf]{revtex}
\tightenlines

\newcommand{\nn}{\nonumber}
\newcommand{\ovl}[1]{\overline{#1}}

\newcommand{\eqn}[1]{(\ref{#1})}

\newcommand{\pslash}{p\kern-1ex /}
\newcommand{\Dslash}{{\cal D}\kern-1.5ex /}
\newcommand{\bpsi}{{\overline{\psi}}}
\newcommand{\bq}{{\overline{q}}}
\newcommand{\tr}{{\rm tr}}
\newcommand{\vev}[1]{\left\langle #1 \right\rangle}

\begin{document}


\title{
\vspace{-3.0cm}
\begin{flushright}  
{\normalsize hep-lat/9906030}\\
{\normalsize UTHEP-405}\\
\end{flushright}
Scaling property of domain-wall QCD in perturbation theory}

\author{ Junichi Noaki and Yusuke Taniguchi}
\address{Institute of Physics, University of Tsukuba, Tsukuba,
 Ibaraki 305-8571, Japan\\}
       
\date{\today}

\maketitle

\begin{abstract}
We estimate the lattice artifacts in loop correction perturbatively for
 domain-wall QCD with infinite number of extra flavors.
We find that there appear no ${\cal O}(a)$ errors in renormalization
 factors of quark wave function, mass and quark bilinear operators
 at one and two loop level
 with off-shell quark momentum.
Our proof is based on even or oddness of the quantum correction in terms 
 of the quark external momentum and mass, and it can be extended to any
 loop level.
\end{abstract}

\pacs{11.15Ha, 11.30Rd, 12.38Bx, 12.38Gc}

\narrowtext

\section{Introduction}

Symanzik's improvement program\cite{Symanzik83} applied to 
on-shell quantities\cite{LuescherWeisz85} 
attempts to eliminate cut-off dependence order by order by 
an expansion in powers of the lattice spacing $a$.  
In order to improve the Wilson quark action to ${\cal O}(a)$ in lattice
QCD, this requires adding the ${\cal O}(a)$ ``clover'' term
\cite{SheikholeslamiWohlert85}. 
Quark operators also have to be modified by ${\cal O}(a)$ counter terms,
which generally involve new operators of higher
dimension\cite{Heatlie91,Jansen96,Luescher9605}. 
In perturbation theory, the tree-level value of the clover coefficient 
and those of the counter terms of quark operators are sufficient to
remove terms of ${\cal O}(g^2a\log a)$ in on-shell Green's functions
evaluated at one-loop level\cite{Heatlie91}.
However there still remains ${\cal O}(g^2a)$ terms and new counter terms 
are needed to eliminate them
\cite{Luescher9606,Sint-Weisz97}.

The domain-wall fermion formulation was originally proposed by
Kaplan\cite{Kaplan92} to define lattice chiral gauge theories with the
introduction of many heavy regulator fields as an extension of the
Wilson fermion.
This original idea and the later work about the Chern-Simons
currents\cite{GoltermanJansenKaplan93} is further developed by
Shamir\cite{Shamir93,Shamir95} into a simpler form and is applied to
lattice QCD (DWQCD) anticipating superior features over other quark
formulations:
no need of the fine tuning to realize the chiral limit
and no restriction for the number of flavors.
Recent simulation results seem to support the former 
feature non-perturbatively\cite{Blum-Soni,Wingate,Blum,Vranas98}. 
It is also perturbatively shown that 
the massless mode at the tree level still remains
stable against the quantum correction\cite{Aoki-Taniguchi}.
In spite of the existence of the Wilson term in the action it does not
directly affect the mass term; there is no additive mass correction.

Related to this good chiral property and disappearance of additive
mass correction from the Wilson term the DWQCD is expected to
have no next leading ${\cal O}(a)$ errors.
It was argued in Ref.~\cite{Wingate} with an intuitive discussion of
good chirality that the leading discretization errors are
${\cal O}(a^2)$ and there is no ${\cal O}(a)$ errors.
They also argue that the scaling behavior of numerical results for
$B_K$\cite{Blum-Soni} and strange quark mass\cite{Wingate} are
consistent with this nature.
For the two dimensional Gross-Neveu model it is reported in 
Ref.\cite{Izb.Nagai99} that the physical observables do not have
${\cal O}(a)$ errors by a nonperturbative analysis of the effective
potential at large $N$ limit.

Although the lattice axial symmetry was discovered\cite{Luscher98}
for the Dirac operator which satisfy the Ginsparg-Wilson
relation\cite{Neuberger,Hasenfratz}
\begin{eqnarray}
\left\{D, \gamma_5\right\} = aDR\gamma_5 D
\end{eqnarray}
and it was shown to exist in the effective theory of DWQCD\cite{KN99},
disappearance of ${\cal O}(a)$ term is not necessarily trivial.
Because the lattice axial transformation of L\"uscher type contains
${\cal O}(a)$ term within the transformation itself,
\begin{eqnarray}
\delta \psi = \gamma_5 (1-aRD) \psi,\quad
\delta \bpsi = \bpsi \gamma_5
\label{eqn:axial-tr}
\end{eqnarray}
and the operators with exact chirality is given by using the projection
\begin{eqnarray}
{\cal O}_\Gamma = \bpsi\Gamma\left(1-\frac{a}{2}RD\right)\psi.
\end{eqnarray}
This projected operator can be written in terms of the boundary fermion
field of the DWQCD ${\cal O}_\Gamma=\bq \Gamma q$\cite{KN99},
where the boundary fields are given by
\begin{eqnarray}
q = \left(1-\frac{a}{2}RD\right)\psi,\quad
\bq = \bpsi
\end{eqnarray}
and are transformed under the axial transformation \eqn{eqn:axial-tr} as
\begin{eqnarray}
\delta q = \gamma_5 q,\quad
\delta \bq = \bq \gamma_5.
\end{eqnarray}
In the context of the DWQCD, where the exact axial symmetry does not
exist because of infinite number of ``unphysical'' fermion fields,
the chirality of the operator is
understood by the axial Ward-Takahashi identity given by Furman and
Shamir with an appropriately defined axial
transformation\cite{Shamir95},
\begin{eqnarray}
\nabla_\mu \vev{A_\mu^a(x) {\cal O}}
=2m \vev{J_5^a(x) {\cal O}}
+\vev{J_{5q}^a(x) {\cal O}}
+i\vev{\delta_A^a {\cal O}},
\end{eqnarray}
where $A_\mu^a$ is an axial current, $J_5^a$ is a pseudo scalar density
and $J_{5q}^a$ is an explicit breaking term.
It was shown in Ref.~\cite{Shamir95} that the exact axial Ward-Takahashi
identity is satisfied for boundary quark operators in the infinite
flavor limit,
\begin{eqnarray}
\nabla_\mu \vev{A_\mu^a(x) {\cal O}(q,\bq)}
=2m \vev{J_5^a(x) {\cal O}(q,\bq)}
+i\vev{\delta_A^a {\cal O}(q,\bq)}.
\end{eqnarray}
The above operator ${\cal O}_\Gamma=\bq \Gamma q$ transforms in the same
way as in the continuum under the transformation and the exact chirality
can be given in order to keep the identity.
Although the axial Ward-Takahashi identity is shown to be realized
nonperturbatively for a well behaved gauge field
configuration\cite{Shamir95},
it is still interesting to see the perturbative understanding of this
good chirality, especially the disappearance of the ${\cal O}(a)$ terms
since we can see the operator mixing structure definitely.

In this paper, we estimate the lattice artifacts in loop correction
perturbatively for DWQCD with infinite width of fifth
dimensional direction.
We show that there appear no ${\cal O}(a)$ errors in renormalization
factors of quark wave function, mass and quark bilinear operators at one
and two loop level.
Although our instrument is perturbation theory
our proof is based on even or oddness of the quantum correction in terms 
of dimensionful quantity such as the quark external momentum and mass, and it 
can be generally extended to any loop level.
We notice that we need not to set the external quark momentum to the
on-shell value to eliminate the ${\cal O}(a)$ errors as in the case of
the Wilson fermion with clover term.
The ${\cal O}(a)$ errors automatically vanish for off-shell quarks like
naively discretized lattice fermion.

This paper is organized as follows.
In Sec.~\ref{sec:model} we introduce the DWQCD action and the Feynman
rules relevant for the present calculation.
In Sec.~\ref{sec:self-energy} we estimate the lattice artifacts for the
quark self energy by expanding the correction in terms of the
external quark momentum and mass.
We start our calculation at one loop level.
We notice that the quantum correction can be classified into odd
function of quark mass and momentum.
Since the relevant term for renormalization is given as a leading order
and the ${\cal O}(a)$ error is a next to leading order term in
expansion, we can see the absence of ${\cal O}(a)$ term quite easily.
This nature of even or oddness is not specific to one loop level we can
apply our procedure to the two loop and any loop level diagrams.
Sec.~\ref{sec:bilinear} is devoted to the estimation of the quark bilinear
operator effective vertex at one and two loop level.
Our conclusion is summarized in Sec.~\ref{sec:concl}.

The physical quantities are expressed in lattice units 
and the lattice spacing $a$ is suppressed unless necessary. 
We take SU($N_c$) gauge group with the gauge coupling $g$.
We set number $N$ of the regulator field or the length of fifth
dimensional direction in domain-wall fermion to infinity.

\section{Action and Feynman rules}
\label{sec:model}

We adopt the Shamir's domain-wall fermion
action\cite{Shamir93},
\begin{eqnarray}
S_{\rm DW} &=&
\sum_{n} \sum_{s=1}^{N} \Biggl[ \frac{1}{2} \sum_\mu
\left( \bpsi(n)_s (-r+\gamma_\mu) U_\mu(n) \psi(n+\mu)_s
+ \bpsi(n)_s (-r-\gamma_\mu) U_\mu^\dagger(n-\mu) \psi(n-\mu)_s \right)
\nn\\&&
+ \frac{1}{2}
\left( \bpsi(n)_s (1+\gamma_5) \psi(n)_{s+1}
+ \bpsi(n)_s (1-\gamma_5) \psi(n)_{s-1} \right)
+ (M-1+4r) \bpsi(n)_s \psi(n)_s \Biggr]
\nn\\&+&
 m \sum_n \left( \bpsi(n)_{N} P_{R} \psi(n)_{1}
+ \bpsi(n)_{1} P_{L} \psi(n)_{N} \right),
\label{eqn:action}
\end{eqnarray}
where $n$ is a four dimensional space-time coordinate and $s$ is an
extra fifth dimensional or ``flavor'' index,
the Dirac ``mass'' $M$ is a parameter of the theory
which we set $0 < M < 2$ to realize the massless fermion at tree
level, $m$ is a physical quark mass,
and the Wilson parameter is set to $r=-1$.
It is important to notice that we have boundaries for the flavor space;
$1 \le s \le N$.
In our one-loop calculation we will take $N\to\infty$ limit and our
proof of vanishing ${\cal O}(a)$ error is valid only in this limit.
$P_{R/L}$ is a projection matrix
\begin{eqnarray}
P_{R/L}=\frac{1\pm\gamma_5}{2}.
\end{eqnarray}
For the gauge part we employ a standard four dimensional
Wilson plaquette action and assume no gauge interaction along the fifth
dimension.

In the DWQCD the zero mode of domain-wall fermion is
extracted by the ``physical'' quark field defined by the boundary
fermions 
\begin{eqnarray}
q(n) = P_R \psi(n)_1 + P_L \psi(n)_{N},
\nn \\
\ovl{q}(n) = \bpsi(n)_{N} P_R + \bpsi(n)_1 P_L.
\label{eq:quark}
\end{eqnarray}
We will consider the QCD operators constructed from this quark fields
and we calculate the lattice artifacts in the Green functions consisting
of only the ``physical'' quark fields.

Weak coupling perturbation theory is developed by writing the link
variable as
$U_{x,\mu}=e^{igA_\mu(x+\hat{\mu}/2)}$
and expanding it in terms of gauge coupling $g$.
The free domain-wall Dirac operator is given as a leading term
\begin{eqnarray}
\Dslash(p)_{st} = \sum_\mu i \gamma_\mu \sin p_\mu \delta_{st}
+\left(W^{+}(p)+mM^{+}\right)_{s,t}P_R
+\left(W^{-}(p)+mM^{-}\right)_{s,t}P_L,
\label{eqn:Dirac-Op}
\end{eqnarray}
where the mass matrix is
\begin{eqnarray}
 W^{+}(p)&=&\left( W^{-}(p) \right)^T=
  \left[
   \begin{array}{cccc}
    -W(p) &   1   &  {}   & {}  \\
    {}    & -W(p) &\ddots & {}  \\
    {}    &  {}   &\ddots &  1  \\
    {}    &  {}   &  {}   &-W(P)
   \end{array}
  \right],
\\
 M^{+}&=&(M^{-})^T=
  \left[
   \begin{array}{ccc}
    {} & {} & {} \\
    1\ \ \ \ & \ {} & \ {} 
   \end{array}
  \right],
\\
 W(p)&=& 1-M-r\sum_\mu(1-\cos p_\mu).
\end{eqnarray}
The domain-wall fermion propagator is given by inverting the Dirac
operator \eqn{eqn:Dirac-Op} for $N\to\infty$ limit\footnote{
Although the $N\to\infty$ limit should be taken after the loop momentum
integral in principle, it is easy to see that the limit can be taken
before the integral by neglecting terms proportional to
$e^{-\alpha N}$.
We take the form of free fermion propagator in the $N\to\infty$ limit
from the beginning.}
\begin{eqnarray}
S_F(p,m)_{st}&=&\left[-i\gamma_\mu\sin p_\mu+W^{-}+mM^{-}\right]_{su}
              G_R(p,m)_{ut}P_R \nonumber \\
           &+&\left[-i\gamma_\mu\sin p_\mu+W^{+}+mM^{+}\right]_{su}
	      G_L(p,m)_{ut}P_L,
	\label{dwfpropagator}
\end{eqnarray}
where sum over the same index is taken implicitly.
$G_{R/L}$ is given by 
\begin{eqnarray}
 G_R(p,m)_{st}&=&G_L(p,m)_{N+1-s,N+1-t}
 \nonumber\\
 &=&-\frac{A(1-m^2)}{F}\left[(1-We^{-\alpha})e^{-\alpha(2N-s-t)}
			+(1-We^{\alpha})e^{-\alpha(s+t)}\right]
 \nonumber\\
 & &-\frac{m}{F}(e^{-\alpha(N-s+t)}+e^{-\alpha(N+s-t)})
  +Ae^{-\alpha|s-t|},
 \\
 \cosh\alpha&=&\frac{1+W^2+\sum_\mu\sin^2 p_\mu}{2|W|},
 \\
 A&=&(2W\sinh\alpha)^{-1},
 \\
 F&=&1-e^{\alpha}W-m^2(1-We^{-\alpha}).
\end{eqnarray}
Note that the argument $p$ of factors $\alpha$ and $W$ is suppressed
in the above formula.
Since we are interested in the Green functions constructed with
``physical'' quark fields the above fermion propagator appears only as
internal quark line.
In order to construct the whole Green function we need other three
types of fermion propagators which connects two ``physical'' quark
fields and ``physical'' quark with fermion field of a general flavor
index,
\begin{eqnarray}
&&
\vev{q(p) \ovl{q}(-p)} = 
 \frac{-i\gamma_\mu \sin p_\mu + \left(1-W e^{-\alpha}\right) m}
{-\left(1-e^{\alpha}W\right) + m^2 (1-W e^{-\alpha})}
 \equiv S_q(p,m),
\label{eqn:phys-prop}
\\&&
\vev{q(p) \bpsi(-p)_s} 
 =\eta_o(p,m)L(p)_s +\eta_e(p,m)R(p)_s,
\label{eqn:q-psi}
\\&&
\vev{\psi(p)_s \ovl{q}(-p)} 
 =R(p)_s\eta_o(p,m) +L(p)_s\eta_e(p,m),
\label{eqn:psi-q}
\end{eqnarray}
where
\begin{eqnarray}
L(p)_s&=&e^{-\alpha(p)(N-s)}P_R+e^{-\alpha(p)(s-1)}P_L,
\\
R(p)_s&=&e^{-\alpha(p)(s-1)}P_R+e^{-\alpha(p)(N-s)}P_L, 
 \label{LRdef}
\\
\eta_o(p,m) &=& \frac{1}{F}
\left(i\gamma_\mu \sin p_\mu - m \left(1 -W e^{-\alpha} \right)\right),
\\
\eta_e(p,m) &=& \frac{e^{-\alpha}}{F} \Bigl[
m \left(i\gamma_\mu \sin p_\mu  -m \left(1-W e^{-\alpha}\right)\right)
- F \Bigr].
\end{eqnarray}
Here we notice that $S_q(p,m)$, $\eta_o(p,m)$ are odd function of
$(p,m)$ and $L(p)_s$, $R(p)_s$, $\eta_e(p,m)$ are even function since 
$W(p)$, $\alpha(p)$ are even function of momentum $p_\mu$,
\begin{eqnarray}
&&
S_q(-p,-m)=-S_q(p,m),\quad
\eta_o(-p,-m)=-\eta_o(p,m),
\\&&
\eta_e(-p,-m)=\eta_e(p,m),\quad
L(-p)_s=L(p)_s,\quad
R(-p)_s=R(p)_s.
\end{eqnarray}
It is important for our later calculation that the quark-fermion
propagator \eqn{eqn:q-psi} and \eqn{eqn:psi-q} can be classified into
even and odd definite part with different ``damping factor''
$L(p)_s$ and $R(p)_s$.

In the perturbative calculation of loop diagrams we take external
momenta and quark masses much smaller than the lattice cut-off, so that
we can expand the external quark propagator in terms of
quark momenta and masses.
In this paper we adopt the following form of expansion in order to
extract the one particle irreducible (1PE) vertex function from the loop
diagram\cite{AIKT98},
\begin{eqnarray}
S_q(p,m) &=& \frac{1-w_0^2}{i\pslash + (1-w_0^2) m}(1+{\cal O}(a^2)),
\label{eqn:qq}
\\
\vev{q(p)\bpsi(-p)_s}&=&\frac{1-w_0^2}{i\pslash+(1-w_0^2)m}
\left[\xi_e(p,m)L(p)_s-\xi_o(p,m)R(p)_s \right],
\label{eqn:qbar}
\\
\vev{\psi(p)_s\bq(-p)}&=&\left[R(p)_s\xi_e(p,m)-L(p)_s\xi_o(p,m)\right]
 \frac{1-w_0^2}{i\pslash+(1-w_0^2)m},
\label{eqn:barq}
\end{eqnarray}
where, $w_o\equiv W(0)=1-M$.
We notice that the quark-fermion propagator \eqn{eqn:qbar} and
\eqn{eqn:barq} are given as multiplication of the free quark propagator
\eqn{eqn:qq} with the linear combination of damping factors, whose
coefficients $\xi_e(p,m),\xi_o(p,m)$ are given in an expanded form,
\begin{eqnarray}
\xi_e(p,m)&=&1+{\cal O}(a^2),
 \nonumber\\
\xi_o(p,m)&=&\frac{w_0}{1-w_0^2}i\pslash+{\cal O}(a^3).
\end{eqnarray}
We call this linear combination as external quark line factor in the
following. 
An explicit form of the next to leading term is not important for our
calculation.
What is relevant for the later discussion is the fact that $\xi_e(p,m)$
and $\xi_o(p,m)$ is an even and odd function in $(p,m)$ and
the even-odd structure of this external line factor is strictly
characterized as a coefficient of damping factors
$L(p)_s$ and $R(p)_s$,
say the even function $\xi_e$ is accompanied with $L_s$ and the odd
function $\xi_o$ with $R_s$ in \eqn{eqn:qbar} and opposite pairing in
\eqn{eqn:barq}. 
We did not expand $L(p)$ and $R(p)$ because their next to leading terms
do not affect the characteristic form of
\eqn{eqn:qbar} \eqn{eqn:barq} but only shift the value of $\xi_{e/o}$
and the damping rate $\alpha$.

Since our gauge part is same as that of the usual Wilson plaquette action,
the gluon propagator can be written as
\begin{eqnarray}
G_{\mu \nu}^{ab} (p)
=\frac{1}{4\sin^2 p/2}
\left[\delta_{\mu \nu}
-(1-\alpha)\frac{4 \sin {p}_\mu/2 \sin {p}_\nu/2}{4 \sin^2 p/2}
\right] \delta_{ab},
\end{eqnarray}
where $\sin^2 p/2 = \sum_\mu \sin^2 p_\mu/2$.
Fermion-gluon vertices are also identical to those in the $N$ flavor
Wilson fermion.
Denoting the interaction vertex with $n$ gluon as $V^{(n)}$ we write
down three of them here,
\begin{eqnarray}
 V^{(1)}(p,k)_{\mu,st}&=&
  -igT^a\left[\gamma_\mu\cos\frac{(-p+k)_\mu}{2}
	 -ir\sin\frac{(-p+k)_\mu}{2}\right] \delta_{st},
\label{eqn:V1}
\\
 V^{(2)}(p,k)_{\mu\nu,st}&=&
  g^2\frac{\{T^a,T^b\}}{2!}\left[i\gamma_\mu
  \sin\frac{(-p+k)_\mu}{2}-r\cos\frac{(-p+k)_\mu}{2}\right]
  \delta_{\mu\nu}\delta_{st},
\label{eqn:V2}
\\
 V^{(3)}(p,k)_{\mu\nu\lambda,st}&=&ig^3\frac{(T^aT^bT^c)_{\rm sym.}}
 {3!}\left[\gamma_\mu\cos\frac{(-p+k)_\mu}{2}-ir\sin\frac{(-p+k)_\mu}{2}
     \right]\delta_{\mu\nu}\delta_{\mu\lambda}\delta_{st},
\label{eqn:V3}
\end{eqnarray}
where $k$ and $p$ represent incoming momentum through the fermion line.
$T^a$ is a generator of color $SU(N_c)$ and $(T^aT^bT^c)_{\rm sym.}$
represents summation over symmetric order of indices $a,b,c$. 

Among several gluon interaction vertices from the pure gauge part we
need an explicit form of the self interaction vertex of three gluons
$A^a_\mu, A^b_\nu, A^c_\lambda$ with incoming momentum $k,l,p$
respectively,
\begin{eqnarray}
G^{(3)}(k,l,p)_{\mu\nu\lambda}
&=&2ig f^{abc} 
\biggl\{
\delta_{\nu \lambda} \sin \frac{1}{2} (p-l)_\mu \cos \frac{k_\nu}{2}
\nn\\&&
+\delta_{\lambda \mu} \sin \frac{1}{2} (k-p)_\nu \cos \frac{l_\lambda}{2}
+\delta_{\mu \nu} \sin \frac{1}{2} (l-k)_\lambda \cos \frac{p_\mu}{2}
\biggr\},
\end{eqnarray}
where $f^{abc}$ is the structure constant of $SU(N_c)$ algebra.
$G^{(3)}$ is an odd function in gluon momentum.

\section{Quark self-energy}
\label{sec:self-energy}

\begin{figure}
\epsfxsize=0.3\textwidth
{\centering\epsfbox{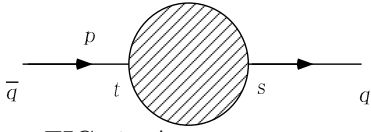} \par}
\caption{A quantum correction to the quark self-energy at some loop
 level . The blob represents the 1PE part.}
\label{any-loop}
\end{figure}

We consider the loop correction to the quark propagator given by a
diagram in Fig.~\ref{any-loop}, where blob represents a quantum correction
of some loop level.
This Green function is written in terms of the 1PE vertex function times 
external quark propagator with help of \eqn{eqn:qbar} \eqn{eqn:barq},
\begin{eqnarray}
\vev{q(p) \ovl{q}(-p)}_1
=
\frac{1-w_0^2}{i\pslash +(1-w_0^2)m}
\Sigma_q(p,m)
\frac{1-w_0^2}{i\pslash +(1-w_0^2)m}.
\label{eqn:one-loop}
\end{eqnarray}
The quark self energy $\Sigma_q(p,m)$ can be expanded in power of 
the external momentum $p$ and mass $m$ keeping the logarithmic dependence 
on them in coefficients, 
\begin{equation}
\Sigma_q(p,m)=\Sigma_0+p_\mu\Sigma_{1\mu}+m\tilde{\Sigma}_1+p^2\Sigma_2
+m^2\tilde{\Sigma}_2+mp_\mu \hat{\Sigma}_{2\mu}+{\cal O}(a^2) ,
\label{eqn:expand}
\end{equation}
where the coefficients $\Sigma_0,\Sigma_{1\mu},\tilde{\Sigma}_1,
\Sigma_{2},\tilde{\Sigma}_2, \hat{\Sigma}_{2\mu}$ are functions of
$\log |p|,\log |m|$ and $p_\mu/m$, 
\begin{equation}
\Sigma_i=\Sigma_i(\log |p|,\log |m|,p/m).
\end{equation}
Here $p_\mu/m$ is introduced for generality.
In \eqn{eqn:expand} the first term represents the additive mass correction,
the second and third term contribute to the quark wave function and
multiplicative mass renormalization factors.
The ${\cal O}(a)$ errors arise from the next three terms. 
The omitted terms give higher order errors of ${\cal O}(a^2)$.
The fact that $\Sigma_0,\Sigma_{1\mu},\tilde{\Sigma}_1,
\Sigma_{2},\tilde{\Sigma}_2$ and $ \hat{\Sigma}_{2\mu}$ are even functions 
of $(p,m)$ is important for our discussion. 
Our viewpoint is that if $\Sigma_q(p,m)$ is an odd function of $(p,m)$ the 
additive mass term and ${\cal O}(a)$ terms do not appear from the first in the 
expansion \eqn{eqn:expand} and only the terms proportional to $p$ or $m$ 
survive to contribute to the corrections of wave function and mass 
multiplicatively. In this case, the leading discretization errors are
${\cal O}(a^2)$.

In the following, we represent the self-energy in the form of loop
integral as
\begin{equation}
 \Sigma_q(p,m)=\int_{l_1 \cdots l_n} \sigma(l_1,\cdots,l_n;p,m),
\end{equation}
where $\int_{l_1 \cdots l_n}$ is the shorthand of $n$ loop integral
\begin{eqnarray}
\int_{l_1 \cdots l_n} 
=\int_{-\pi}^{\pi}\frac{d^4l_1}{(2\pi)^4}\cdots\frac{d^4l_n}{(2\pi)^4},
\end{eqnarray}
 which is invariant against the sign flip of $l_1,\cdots,l_n$.
We concentrate to show the oddness of $\sigma(l_1,\cdots,l_n;p,m)$ for the
variables $l_1,\cdots,l_n,p,m$.
If it is valid, $\Sigma_q(p,m)$ becomes odd function of $(p,m)$:
\begin{eqnarray}
\Sigma_q(-p,-m)&=&\int_{l_1,\cdots,l_n}\sigma(l_1,\cdots,l_n;-p,-m) \nonumber\\
 	&=&\int_{l_1,\cdots,l_n}\sigma(-l_1,\cdots,-l_n;-p,-m) \nonumber \\
	&=&-\Sigma_q(p,m).
\end{eqnarray}

\begin{figure}
\epsfxsize=0.8\textwidth
{\centering\epsfbox{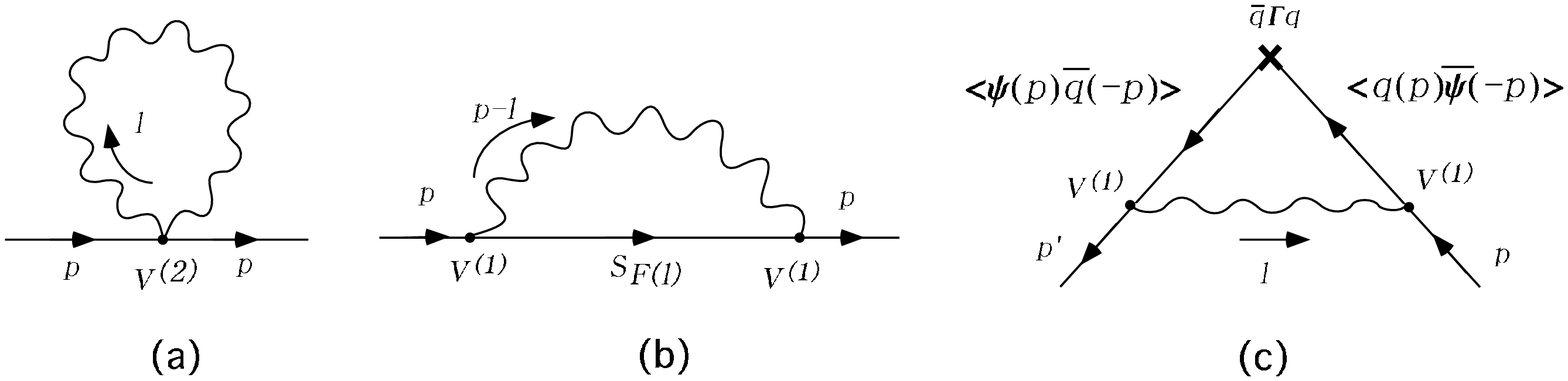} \par}
\caption{The one loop contributions to the quark self-energy;(a),(b) and 
the quark bilinear operator;(c).}
\label{1-loop}
\end{figure}

We begin by considering the one loop contributions,
 which are given by ``tadpole'' and ``half-circle'' diagrams 
of Fig.~\ref{1-loop} (a) and (b) respectively.
As was discussed in the above $\sigma(l;p,m)$ is given by multiplying
the self energy of domain-wall fermion by the external quark line
factors $[\xi_eL_s-\xi_oR_s]$, $[R_s\xi_e-L_s\xi_o]$
of the propagator \eqn{eqn:qbar} \eqn{eqn:barq}.
Each diagram gives the following integrand,
\begin{eqnarray}
 \sigma^{\rm tadpole}(l;p,m)&=&[\xi_e L_s-\xi_o R_s](p,m)V^{(2)}(-p,p)_
  {\mu\nu,st}[R_t\xi_e-L_t\xi_o](p,m)G_{\mu\nu}(l),
\label{tadp}
\\
 \sigma^{\rm half-circle}(l;p,m)&=&[\xi_e L_s-\xi_o R_s](p,m)
 V^{(1)}(-p,l)_{\mu,st}S_F(l,m)_{tu}V^{(1)}(-l,p)_{\nu,uv} G_{\mu\nu}(p-l)
 \nn\\&\times&
 [R_v\xi_e-L_v\xi_o](p,m).
\label{halfc}
\end{eqnarray}
We take summation over flavor index step by step from the left 
paying attention to the external quark line factor
$[\xi_e L_s-\xi_o R_s](p,m)$ .
A characteristic property of the factor is that an even and odd function
is separated by the damping factor $L(p)_s$ and $R(p)_s$.
We can see that this separation property is preserved under each steps
in the diagram.

By using the commutation relation
\begin{eqnarray}
L(p)_s\gamma_\mu&=&\gamma_\mu R(p)_s
\label{commute}
\end{eqnarray}
and the form of internal fermion propagator in \eqn{dwfpropagator}, the
multiplication of the damping factor with the fermion propagator
becomes,
\begin{eqnarray}
 L(p)_sS_F(l,m)_{st}&=&f_eL(p)_t+h_eL(l)_t+f_oR(p)_t+h_oR(l)_t,
  \label{eqn:LS}
 \\
 R(p)_sS_F(l,m)_{st}&=&f'_eR(p)_t+h'_eR(l)_t+f_oL(p)_t+h'_oL(l)_t,
  \label{eqn:RS}
\end{eqnarray}	 
where
\begin{eqnarray}
f_o &=& i\gamma_\mu\sin l_\mu X,
\\
f_e &=& (W(l)-e^{\alpha(p)})X,
\\
h_o &=& i\gamma_\mu \sin l_\mu Y_2
 +\frac{m(W(l)-e^{\alpha(p)})}{F(l)(e^{\alpha(l)}-e^{-\alpha(p)})}
 +\frac{me^{\alpha(p)-\alpha(l)}}{F(l)},
\\
h_e &=& \frac{im\gamma_\mu \sin l_\mu}{F(l)(e^{\alpha(l)}-e^{-\alpha(p)})}
 +(W(l)-e^{\alpha(p)})Y_1+\frac{e^{\alpha(p)}}{F(l)},
\\
f'_e &=& (W(l)-e^{-\alpha(p)})X,
\\
h'_o &=& i\gamma_\mu \sin l_\mu Y_1
 +\frac{m(W(l)-e^{-\alpha(p)})}{F(l)(e^{\alpha(l)}-e^{-\alpha(p)})}
 -\frac{m}{F(l)},
\\	
h'_e &=& \frac{im\gamma_\mu \sin l_\mu}{F(l)(e^{\alpha(l)}-e^{-\alpha(p)})}
 +(W(l)-e^{-\alpha(p)})Y_2-\frac{m^2e^{-\alpha(l)}}{F(l)},
\\
X^{-1}&=&W(l)(e^{\alpha(p)}+e^{-\alpha(p)}-e^{\alpha(l)} -e^{-\alpha(l)}),
\\
Y_1&=&A(l)\left[\frac{(1-m^2)(1-We^{-\alpha(l)})}
	   {F(l)(1-e^{-\alpha(p)-\alpha(l)})}
	   -\frac{1}{1-e^{-\alpha(p)+\alpha(l)}}\right],
\\
Y_2&=&A(l)\left[\frac{(1-m^2)(1-We^{\alpha(l)})e^{-2\alpha(l)}}
	   {F(l)(1-e^{-\alpha(p)-\alpha(l)})}
	   -\frac{1}{1-e^{-\alpha(p)+\alpha(l)}}\right].	
\end{eqnarray} 	
Among several coefficients in \eqn{eqn:LS} and \eqn{eqn:RS},
$f_e,\ h_e$ and $h'_e$ are even functions and $f_o,\ h_o$ and $h'_o$ are
odd.
Multiplication of the external line factor with the internal quark line
becomes
\begin{eqnarray}
 [\xi_e L_s-\xi_o R_s](p)S_F(l,m)_{st}&=&(\xi_ef_e-\xi_of_o)L(p)_t
  +(\xi_e h_e-\xi_o h'_o)L(l)_t
\nonumber\\
 &+&(\xi_ef_o-\xi_of'_e)R(p)_s+(\xi_eh_o-\xi_oh'_e)R(l)_t.
\label{dampSF}
\end{eqnarray}
It should be noticed that the even-odd structure in which $ L_s$'s have even 
coefficients and $R_s$'s have odd ones is preserved.

Multiplication with the interaction vertex $V^{(n)}$ is given by
\begin{eqnarray}
 &&[\xi_eL_s-\xi_oR_s](p)V^{(1)}(p,k)_{\mu,st}=-ig
  (u_{\mu o}L_t+u_{\mu e}R_t)(p),
\label{dampV1}
\\
 && \left[\xi_eL_s-\xi_oR_s\right](p)V^{(2)}(p,k)_{\mu\nu,st}=
  g^2(u_{\mu e}L_t+u_{\mu o}R_t)(p)\delta_{\mu\nu},
\label{dampV2}
\\
 && \left[\xi_eL_s-\xi_oR_s\right](p)V^{(3)}(p,k)_{\mu\nu\lambda,st}=
  ig^3(u_{\mu o}L_t+u_{\mu e}R_t)(p)\delta_{\mu\nu}\delta_{\mu\lambda},
\label{dampV3} 
\end{eqnarray}
where $u_{\mu e}$ and $u_{\mu o}$ are even and odd function,
\begin{eqnarray}
 u_{\mu e}(p,k,m)&=&\gamma_\mu\xi_e(p,m)\cos\frac{(-p+k)_\mu}{2}
  +ir\xi_o(p,m)\sin\frac{(-p+k)_\mu}{2},
\\
 u_{\mu o}(p,k,m)&=&\gamma_\mu\xi_o(p,m)\cos\frac{(-p+k)_\mu}{2}
  +ir\xi_e(p,m)\sin\frac{(-p+k)_\mu}{2},
\end{eqnarray}
and we omitted the color factors. In general, the interaction vertex 
with even number of gluons $V^{(2n)}$ preserves
the structure of damping factor and that with odd gluons $V^{(2n+1)}$
flips it so as to assigning odd coefficients to $L_s$'s and even to
$R_s$'s.

After several multiplication in the diagram the external line factor
from the left meets the other factor from the right and flavor
index is summed over using the formula,
\begin{eqnarray}
L(p_1)_sL(p_2)_s &=& R(p_1)_sR(p_2)_s
 = \frac{1}{1-e^{-\alpha(p_1)-\alpha(p_2)}}
 \equiv C(p_1,p_2),
\nonumber
\\
L(p_1)_sR(p_2)_s &=&0.
\label{ortho} 
\end{eqnarray}

We apply this scenario $\sigma^{\rm tadpole}(l;p,m)$.
The left factor $[\xi_e L_s-\xi_o R_s](p,m)$ is multiplied to $V^{(2)}$
but this does not change the even-odd structure according to
\eqn{dampV2} and then meets the right factor and becomes an odd
function,
\begin{eqnarray}
[\xi_e' L_s-\xi_o' R_s](p,m)[R_t\xi_e-L_t\xi_o](p,m)
=-C(p,p)\left(\xi_e'\xi_o+\xi_o'\xi_e\right).
\end{eqnarray}
In $\sigma^{\rm half-circle}(l;p,m)$, the 
flip of the even-odd structure occurs twice at each interaction vertex 
$V^{(1)}$ with \eqn{dampV1}.
Multiplication with the fermion propagator does not change the structure
as was shown in \eqn{dampSF}.
Finally the left factor meets the right factor keeping the original
structure and gives an odd function,
\begin{eqnarray}
[\xi_e^{''} L_s-\xi_o^{''} R_s](p,m)[R_t\xi_e-L_t\xi_o](p,m)
=-C(p,p)\left(\xi_e^{''}\xi_o+\xi_o^{''}\xi_e\right).
\end{eqnarray}
As a result, it is shown that $\sigma(l;p,m)$ is an odd function of $(l,p,m)$ 
at one loop level together with the fact that gluon propagator is an even 
function. Consequently the additive mass correction and ${\cal O}(a)$ errors
vanish.

\begin{figure}
\epsfxsize=0.8\textwidth
{\centering\epsfbox{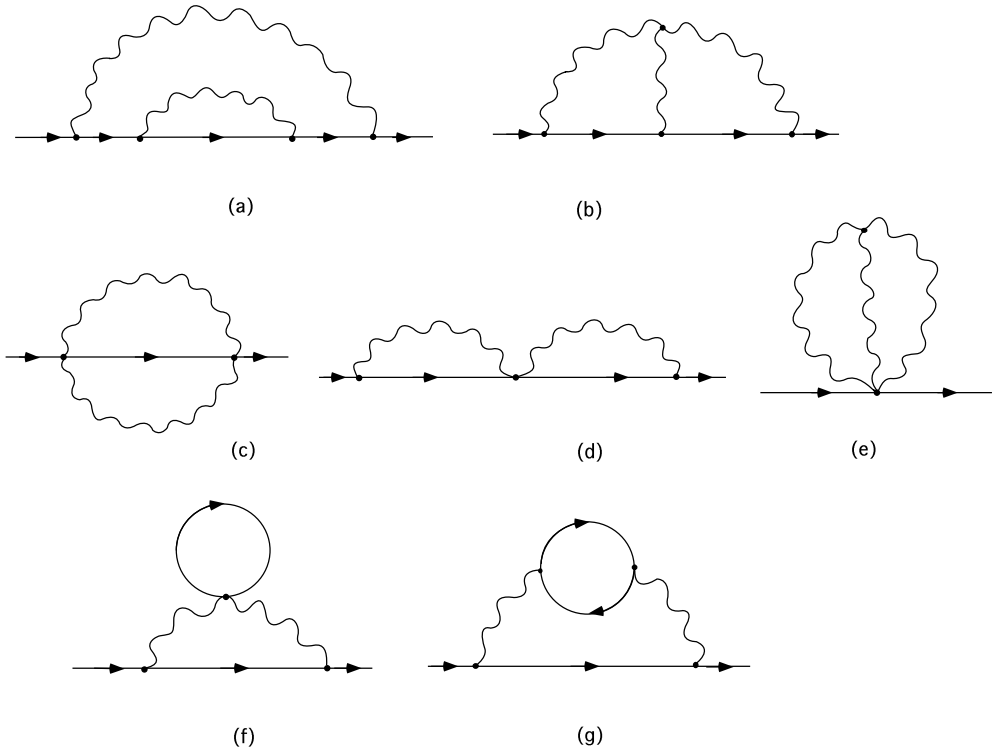} \par}
\caption{Typical contributions to the quark self-energy at two loop level.}
\label{2-loop_self}
\end{figure}

It is straightforward to extend the above discussion to two loop level.
As examples we consider the diagrams of Fig.~\ref{2-loop_self} (a), (b),
whose integrands are
\begin{eqnarray}
\sigma^{(a)}(l_1,l_2;p,m) &=& [\xi_e L_s-\xi_o R_s](p,m)
\nn\\&\times&
 V^{(1)}(-p,l_1)_\mu S_F(l_1)_{st}
 V^{(1)}(-l_1,l_2)_\rho S_F(l_2,m)_{tu}
 V^{(1)}(-l_2,l_1)_\sigma S_F(l_1,m)_{uv}
\nn\\&\times&
 V^{(1)}(-l_1,p)_\nu 
G_{\mu\nu}(p-l_1) G_{\rho\sigma}(l_1-l_2)
[R_v\xi_e-L_v\xi_o](p,m),
\\
\sigma^{(b)}(l_1,l_2;p,m) &=&[\xi_e L_s-\xi_o R_s](p,m)
\nn\\&\times&
 V^{(1)}(-p,l_2)_\mu S_F(l_2,m)_{st}
 V^{(1)}(-l_2,l_1)_\nu S_F(l_1,m)_{tu}
 V^{(1)}(-l_1,p)_\rho
\nn\\&\times&
 G^{(3)}(-p+l_2,l_1-l_2,p-l_1)_{\alpha\beta\gamma}
 G_{\mu\alpha}(p-l_2) G_{\nu\beta}(l_1-l_2) G_{\rho\gamma}(p-l_1)
\nn\\&\times&
[R_v\xi_e-L_v\xi_o](p,m).
\end{eqnarray}
There are four interaction vertices $V^{(1)}$ in $\sigma^{(a)}(l_1,l_2;p,m)$.
Since the fermion propagator does not affect the even-odd argument,
the even-odd structure of the left factor
$[\xi_e L_s-\xi_o R_s](p,m)$ is flipped four times and finally meets
with the right factor in the same form as in the tree level
\begin{eqnarray}
[\xi_e' L_s-\xi_o' R_s][R_v\xi_e-L_v\xi_o]
\end{eqnarray}
and gives an odd contribution.
Combined with the gluon propagator that is an even function, we can see
$\sigma^{(a)}(l_1,l_2;p,m)$ is odd.
On the other hand, there are three interaction vertices $V^{(1)}$ in
$\sigma^{(b)}(l_1,l_2;p,m)$.
This flips the even-odd structure of the left factor three times and the 
external quark line factor gives even contribution,
\begin{eqnarray}
[\xi_o^{''} L_s-\xi_e^{''} R_s][R_v\xi_e-L_v\xi_o].
\end{eqnarray}
However, diagram $\sigma^{(b)}(l_1,l_2;p,m)$ also contains the three gluon 
self interaction vertex $G^{(3)}$ which is an odd function of $(l_1,l_2,p)$.
Multiplying this term, the total loop integrand is an odd function.
Confirmation of oddness for other remaining two loop diagrams is
straightforward with the help of \eqn{dampSF}-\eqn{dampV3}.
We leave it to reader and just pick up three of them in
Fig.~\ref{2-loop_self} (c), (d) and (e) as examples.
The above procedure is applicable to any quark self-energy diagram at
any loop level and shows the oddness of the loop correction.
The ${\cal O}(a)$ errors vanish for any diagrams.

In the end we have two comments on two loop diagrams
Fig.~\ref{2-loop_self} (f), (g) with fermion loop in the gluon
polarization.
The loop correction to the gluon polarization does not spoil the evenness
of the gluon propagator because it is protected by gauge symmetry.
Although it is almost trivial for the DWQCD from the same reason, 
we checked it explicitly at one loop level.
This is accomplished by evaluating
\begin{eqnarray}
I_{\mu\nu}(l,p) &=&
-\tr\left[
 V^{(1)}(-l,l+p)_\mu S_F(l+p,m)_{st} V^{(1)}(-l-p,l)_\nu S_F(l,m)_{ts}
 \right]
\nn\\&&
-\tr\left[V^{(2)}(-l,l)_{\mu\nu} S_F(l,m)_{ss}\right].
\end{eqnarray}
Since the evenness is seen easily by substituting
\eqn{dwfpropagator}, \eqn{eqn:V1} and \eqn{eqn:V2}, we omit the
explicit calculation. The oddness of the three gluon self interaction 
vertex is also protected by gauge symmetry.
The second comment is concerning the Pauli-Villars field needed to
settle the infra-red divergence in fermion loop with infinite number of
domain-wall fermion.
The Pauli-Villars field is introduced as the $N$ flavor Wilson Dirac
boson whose Dirac operator is the same as that of the domain-wall fermion
except for the physical quark mass is changed to the cut-off order and
opposite signature; $m\to-1$ \cite{Vranas97} and it keeps the gauge
invariance.
This does not change the even-odd structure of the Feynman rules in pure
gauge part.

\section{Quark bilinear operator}
\label{sec:bilinear}

We consider quark bilinear operators in the following form
\begin{eqnarray}
{\cal O}_\Gamma(x) = \ovl{q}(x) \Gamma q(x)
,\qquad
\Gamma= 1,\,\gamma_5,\,\gamma_\mu,\,\gamma_\mu\gamma_5,\,\sigma_{\mu \nu}.
\label{eq:bilinear}
\end{eqnarray}
We calculate the quantum correction to the Green's function
$\vev{{\cal O}_\Gamma(x) q(y) \ovl{q}(z)}$ with external momenta
$p, p^\prime$,
As in the previous section we see that the external line is
essentially written in terms of the quark propagator times damping
factors in \eqn{eqn:qbar} and \eqn{eqn:barq}.
Making use of this fact the full Green's function for small external
momentum becomes
\begin{eqnarray}
\vev{(\ovl{q} \Gamma q) \cdot q \ovl{q}}_{\rm full}
=
\frac{\left(1-w_0^2\right) Z_w Z_2}{i\pslash'+(1-w_0^2)Z_w Z_m^{-1}m}
 T_\Gamma(p',p,m) \Gamma
\frac{\left(1-w_0^2\right) Z_w Z_2}{i\pslash+(1-w_0^2)Z_w Z_m^{-1}m},
\end{eqnarray}
where $T_\Gamma$ is the loop correction to the operator vertex and
$Z_2$, $Z_w$, $Z_m$ are quantum correction to quark wave function, quark
over all factor $w_0$ and mass whose explicit value is given in
Ref.~\cite{AIKT98} at one loop.
From the result of previous section, it is proven that the 
${\cal O}(a)$ errors vanish automatically for $Z_2$, $Z_w$, $Z_m$.
In this section we show that for $T_\Gamma$.

As in the previous section
we expand the effective vertex $T_\Gamma(p,p',m)$ in terms of 
external quark momentum $p,p'$ and mass $m$ keeping the logarithmic 
dependence on them in coefficients, and then estimate the lattice 
artifact:
\begin{equation}
 T_\Gamma(p',p,m)= T_0+p_\mu T_{1\mu}+p'_\mu T'_{1\mu}+m \tilde{T}_1+
{\cal O}(a^2).
\end{equation}
The coefficients $T_0,T_1$ and $T'_1$ are functions of 
$\log |p|,\log |p'|,\log |m|,p/m,
p'/m$. $T_0$ contributes to the renormalization factor of 
the bilinear operator and following three terms are ${\cal O}(a)$ errors.
If $T_\Gamma(p',p,m)$ is an even function of $(p',p,m)$, the
${\cal O}(a)$ errors vanish automatically.
In the following we adopt the loop integral form as
\begin{equation}
 T_\Gamma(p',p,m)=\int_{l_1 \cdots l_n} \gamma(l_1,\cdots,l_n;p',p,m).
\end{equation}
$T_\Gamma$ is even provided the integrand $\gamma(l_1,\cdots,l_n;p',p,m)$ 
is even function of $(l_1,\cdots,l_n,p,p',m)$.

At one loop level, $\gamma(l;p',p,m)$ 
is given by a diagram of Fig.~\ref{1-loop}(c) as
\begin{eqnarray}
\gamma(l;p',p,m)
&=&[\xi_e L_s-\xi_o R_s](p',m)V^{(1)}(-p',p'+l)
[R_s\eta_o+L_s\eta_e](p'+l,m)\Gamma \nonumber\\
&\times& [\eta_oL_t+\eta_eR_t](p+l,m)V^{(1)}(-p-l,p)
  [R_t\xi_e-L_t\xi_o](p,m)G_{\mu\nu}(l),
 \label{eqn:tq}
\end{eqnarray} 
where the quark propagators \eqn{eqn:q-psi} and \eqn{eqn:psi-q} are used 
as internal lines, since ``physical'' fields are living in the 
operator vertex.
Each internal line propagators contain damping factors $L_s$, $R_s$
and consideration of even-odd structure is carried for left and
right hand side of fermion line independently.
There is one fermion interaction vertex $V^{(1)}$ in each fermion
lines of \eqn{eqn:tq} and flip of even-odd structure occurs once.
This gives odd functions,
\begin{eqnarray}
[\xi_o' L_s-\xi_e' R_s][R_s\eta_o+L_s\eta_e]
\end{eqnarray}
for the left side line and
\begin{eqnarray}
[\eta_e' L_t+\eta_o' R_t][R_t\xi_e-L_t\eta_o]
\end{eqnarray}
for right side one making the total contribution even.

\begin{figure}
\epsfxsize=0.8\textwidth
{\centering\epsfbox{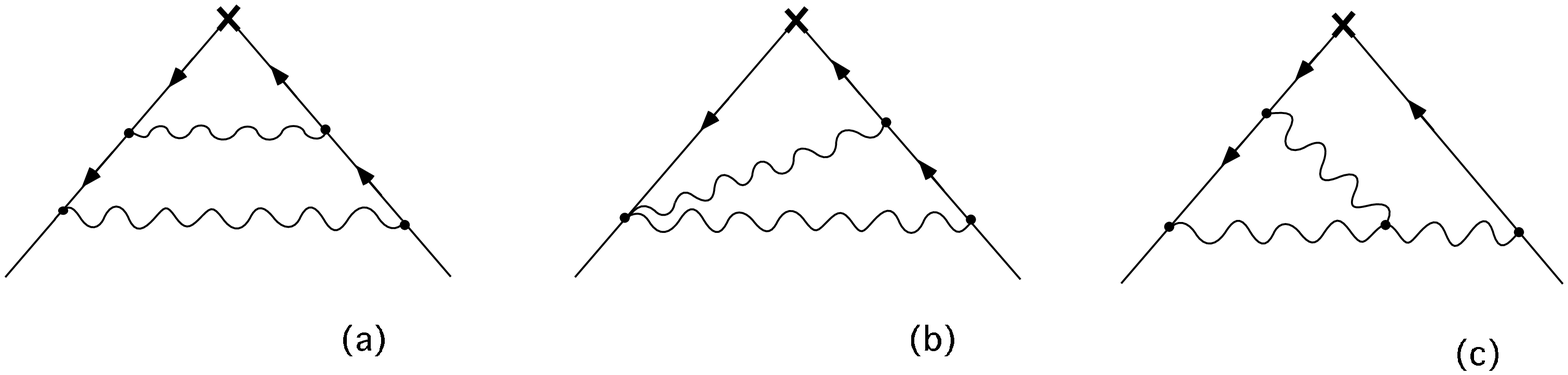} \par}
\caption{Typical contributions to the quark bilinear operator at
 two loop level.}
\label{2-loop_cur}
\end{figure}

For two loop diagrams, we consider Fig.~\ref{2-loop_cur}(a) as an example.
There are two interaction vertices $V^{(1)}$ in each fermion line, 
where the even-odd flip occurs twice giving odd contribution.
Therefore, the total contribution from the multiplication of them becomes
even as expected.
The above argument is valid for all other remaining two loop diagrams.
For example estimation of  diagrams in Fig.~\ref{2-loop_cur}(b),(c) 
is an easy task and we leave it to reader.

\section{Conclusion}
\label{sec:concl}

In this article, we made an estimation of the lattice artifacts in loop
correction perturbatively for DWQCD in $N\to\infty$ limit.
We studied quantum corrections to the Green function constructed from
the physical quark field only.
We revealed the even-odd structure of the diagrams
in terms of the external quark momentum and mass.

With the help of the damping factors in the external quark propagator,
we have shown to two loop level that 
the quark self-energy is definitely odd and 
the effective vertex of the quark bilinear operator is even.
This means that the quark self-energy has neither additive mass correction 
nor ${\cal O}(a)$ errors, and ${\cal O}(a)$ errors in the effective vertex
of the quark bilinear operator vanish automatically.
Since our discussion can be extended to any loop diagram,
we can show that ${\cal O}(a)$ errors do not appear for any diagram 
at any loop level in the same way.
This is because the disappearance of them is deeply concerned with 
the hidden axial symmetry\cite{Luscher98}  of the DWQCD which was
explicitly shown to exist in Ref.~\cite{KN99} for effective Lagrangian
of the physical quark fields.
       
\section*{Acknowledgments}
We thank S. Aoki for important suggestions and reading the manuscript. 
We also appreciate A. Ukawa, Y. Kikukawa, S. Ejiri, T. Izubuchi and K. Nagai 
for valuable comments and discussions. A part of this work is supported by
the Grants-in-Aid for Scientific Research from the Ministry of Education, 
Science and Culture (No.2373).  Y.~T. is supported by Japan Society
for Promotion of Science.

\newcommand{\J}[4]{{#1} {\bf #2} (19#3) #4}
\newcommand{\MPL}{Mod.~Phys.~Lett.}
\newcommand{\IJMP}{Int.~J.~Mod.~Phys.}
\newcommand{\NP}{Nucl.~Phys.}
\newcommand{\PL}{Phys.~Lett.}
\newcommand{\PR}{Phys.~Rev.}
\newcommand{\PRL}{Phys.~Rev.~Lett.}
\newcommand{\AP}{Ann.~Phys.}
\newcommand{\CMP}{Commun.~Math.~Phys.}
\newcommand{\PTP}{Prog. Theor. Phys.}
\newcommand{\Suppl}{Prog. Theor. Phys. Suppl.}


\end{document}